\documentclass[11pt,a4paper]{article}

\usepackage[T1]{fontenc}
\usepackage[utf8]{inputenc} 
\usepackage{lmodern}  
\usepackage{geometry}
\usepackage{microtype} 
\usepackage{setspace}  
\usepackage{parskip}   
\usepackage{listings}
\usepackage[table]{xcolor}
\usepackage{tcolorbox}

\usepackage{colortbl}
\usepackage{array}
\usepackage{caption}
\tcbuselibrary{listings}
\usepackage[authoryear]{natbib}
\definecolor{stringcolor}{rgb}{0.4,0.7,0.4}
\definecolor{numbercolor}{rgb}{0.5,0.5,1}
\definecolor{keycolor}{rgb}{0.1,0.4,0.7}

\lstdefinestyle{jsonstyle}{
    language=Java, 
    basicstyle=\small\ttfamily,
    morestring=[b]",
    moredelim=[il][\color{keycolor}]{-}, 
    stringstyle=\color{stringcolor},
    keywordstyle=\color{numbercolor},
    morekeywords={true,false,null},
    showstringspaces=false,
    breaklines=true,
    frame=none
}

\usepackage{graphicx}
\usepackage{float}

\usepackage{amsmath,amssymb,amsthm}

\theoremstyle{remark}

\usepackage{booktabs} 
\usepackage{siunitx}   
\usepackage{hyperref}
\hypersetup{
  colorlinks=true,
  linkcolor=black,
  citecolor=black,
  urlcolor=black,
  pdfauthor={Raunak Narwal},
  pdftitle={BIGBOY1.2}
}
\usepackage[capitalize,nameinlink]{cleveref} 

\usepackage{authblk}

\newif\ifTitleSmallCaps
\TitleSmallCapstrue 

\newcommand{\TitleCase}[1]{\ifTitleSmallCaps\textsc{#1}\else#1\fi}

\newcommand{\headertext}{}

\newcommand{\shorttitle}{BIGBOY1.2}

\definecolor{lightgray}{gray}{0.95} 
\definecolor{midgray}{gray}{0.5} 

\usepackage{fancyhdr}
\newcommand{\keywords}[1]{\par\vspace{0.5em}\noindent\textbf{Keywords:}~#1\par}

\renewenvironment{abstract}{%
  \vspace{1em}%
  \begin{center}\bfseries\large\scshape Abstract\end{center}%
  \begin{quote}
}{\end{quote}\vspace{1em}}

\usepackage{titlesec}
\titlespacing*{\section}{0pt}{2.0ex plus 0.5ex minus 0.2ex}{1.0ex}
\titlespacing*{\subsection}{0pt}{1.5ex plus 0.5ex minus 0.2ex}{0.8ex}
\titlespacing*{\subsubsection}{0pt}{1.0ex plus 0.3ex minus 0.2ex}{0.5ex}

\titleformat{\section}{\large\bfseries}{\thesection}{1ex}{}
\titleformat{\subsection}{\normalsize\bfseries}{\thesubsection}{1ex}{}
\titleformat{\subsubsection}{\normalsize\itshape}{\thesubsubsection}{1ex}{}

\title{\TitleCase{BIGBOY1.2: \textsc{Generating Realistic Synthetic Data for Disease Outbreak Modelling and Analytics}}}
\author[1]{Raunak Narwal\thanks{\texttt{ms23177@iisermohali.ac.in}}}
\author[2]{Syed Abbas\thanks{\texttt{Corresponding author: sabbas.iitk@gmail.com, abbas@iitmandi.ac.in}}}
\affil[1]{Department of Mathematical Sciences, Indian Institute of Science Education and Research, Mohali, 140306, Punjab, India}
\affil[2]{School of Mathematical and Statistical Sciences \\
Indian Institute of Technology Mandi, Mandi, 175005, H.P., India}
\date{\today}
\begin{document}

\maketitle

\begin{abstract}
Modelling disease outbreak models remains challenging due to incomplete surveillance data, noise, and limited access to standardized datasets. We have created \textbf{BIGBOY1.2}, an open synthetic dataset generator that creates configurable epidemic time series and population-level trajectories suitable for benchmarking modelling, forecasting, and visualisation. The framework supports SEIR and SIR-like compartmental logic, custom seasonality, and noise injection to mimic real reporting artifacts. BIGBOY1.2 can produce datasets with diverse characteristics, making it suitable for comparing traditional epidemiological models (e.g., SIR, SEIR) with modern machine learning approaches (e.g., SVM, neural networks). \par
\end{abstract}

\section{Introduction}
Infectious diseases have repeatedly challenged the global health system, economies, and societies. During the past few decades, we have witnessed outbreaks such as SARS (2003)\cite{who_sars}, Ebola\cite{who_ebola}, and COVID-19\cite{who_covid}, which have shown how quickly pathogens can disrupt normal life and healthcare systems, causing unprecedented economic and social consequences. In such scenarios, epidemic modeling plays a significant role in disease outbreak prediction, policymaking, and timely intervention strategies\cite{keeling_rohani}. But epidemiological modelling remains heavily dependent on the availability and quality of outbreak data. Missing dates, reporting delays, and substandard datasets make it challenging to train and benchmark models. That has led to our increased interest in synthetic data generation, BIGBOY1.2 could generate reality mimicking datasets and visualizations which make them ideal for benchmarking models, stress testing algorithms, and conducting reproducible experiments.

\subsection{Background}
Accurate modeling and forecasting of disease outbreaks have been a crucial topic for public health planning and decision making. Classical epidemiological models, such as compartmental models (SIR, SEIR)\cite{hethcote}, have proven their effectiveness for understanding transmission dynamics. These can be used to estimate parameters like basic reproduction number $R_0$, beta effective, and estimate interventions. However, these models are idealistic and generally different from real world data. Real world data is noisy, incomplete, and subject to irregular reporting due to many factors such as delays in case confirmation, underreporting, and inconsistent testing policies across different regions\cite{holmdahl_buckee}. \\
The COVID-19 pandemic further highlighted the need for high quality datasets for epidemic modelling\cite{chinazzi_covid}. Most studies relied on the use of fragmented and incomplete datasets, which limited the reliability of forecasts and their ability to reproduce. Data inconsistencies such as negative incidence values (due to backlogs and correction) created major challenges for data-driven machine learning models, that require large, well structured datasets. As a result, forecasting methods on real world datasets are often inconclusive. \\
To limitations have compelled researchers to use synthetic data. Our synthetic data generator , BIGBOY1.2 allows for complete control over epidemic parameters, which includes population, layers, seasonality, interventions, and stochastic variations. It provides an invaluable testbed for benchmarking forecasting models under controlled scenarios, enabling rigorous evaluation of algorithmic performance in conditions where real world data would be insufficient or biased. But most existing synthetic dataset tools are either very simplistic and fail to mimic the complex nature of real world outbreaks or too specialized, designed for specific diseases and narrow research goals\cite{venkatramanan2020}.

\subsection{Motivation for BIGBOY1.2}
As discussed before, synthetic data generators fall short in key aspects of realism, flexibility and usability. Important factors like seasonal variation, stochastic effects and reporting biases are ignored and are tightly coupled to specific diseases or parameters settings. As a consequence, researchers resort to creating ad-hoc datasets, which lack standardization, making it difficult to compare forecasting models across studies\cite{ray2020}. \\
Moreover, current tools rarely integrate visual analytics with the data generation pipeline. The ability to intuitively visualize compartmental dynamics and intervention impacts is very crucial for communicating findings effectively. Without built-in visualization support, the user relies on external scripts and tools, increasing complexity to even perform a basic exploratory analysis. \\
We have proposed BIGBOY1.2 , a versatile and fully configurable synthetic dataset generator for disease outbreak modeling and analytics. BIGBOY1.2 allows users to simulate epidemics with customizable transmission parameters and intervention strategies. It also generates visual plots such as time series plots, heatmaps, phase diagrams along with datasets. BIGBOY1.2 is lightweight and easy to use, unlike many heavy ML-based dataset generators. By standardizing synthetic dataset creation, BIGBOY1.2 aims to improve reproducibility and enable fair benchmarking of disease outbreak modeling.

\section{Methods}
BIGBOY1.2 is a stochastic epidemic dataset and visual plot generator which builds upon BIGBOY1. With further refinements , it presents better and more realistic datasets.

\subsection{Framework}
BIGBOY1.2 is designed to simulate realistic infectious disease outbreaks with a high degree of configurability and realism. At its core, it is built on the SEIR (Susceptible, Exposed, Infectious, Recovered) model\cite{hethcote}, extended with dynamic parameters, seasonal influences, vaccination, and multi-wave outbreak structures\cite{bakhta2022}. It is a multi-layered and age-structured SEIR model which includes a noise and reporting module to simulate real-world data irregularities\cite{gostic2020}. Unlike traditional simulations, BIGBOY1.2 produces data that closely resembles real-world epidemic curves and also retains full control over the underlying "ground truth" parameters. This allows researchers to test forecasting methods under controlled conditions. \\
the framework is modular in design, it consists of four key layers: Parameter Initialization, where user can manually define epidemiological and behavioral parameters; Simulation Engine, integrates the SEIR based equations and accounts for time varying transmission dynamics, interventions and stochastic effects; Noise and Reporting Layers, injects realistic data artifacts like under reporting, reporting delays and random fluctuations to mimic real world surveillance data and the last layer is Output and Visualizations, which exports the datasets in csv formats, parameters in JSON format and generates a range of visual graphs from simple time series plots to advanced 3D plots\cite{venkatramanan2020}. \\
BIGBOY1.2 supports three operational modes through CLI : random mode (parameters taken from predefined ranges), interactive mode (user-driven configuration), and batch mode (generates many simulations at once). The user could also use various commands from the CLI like --plots all , --population X.\\

\begin{figure}[H] 
    \centering
    \includegraphics[width=0.8\linewidth]{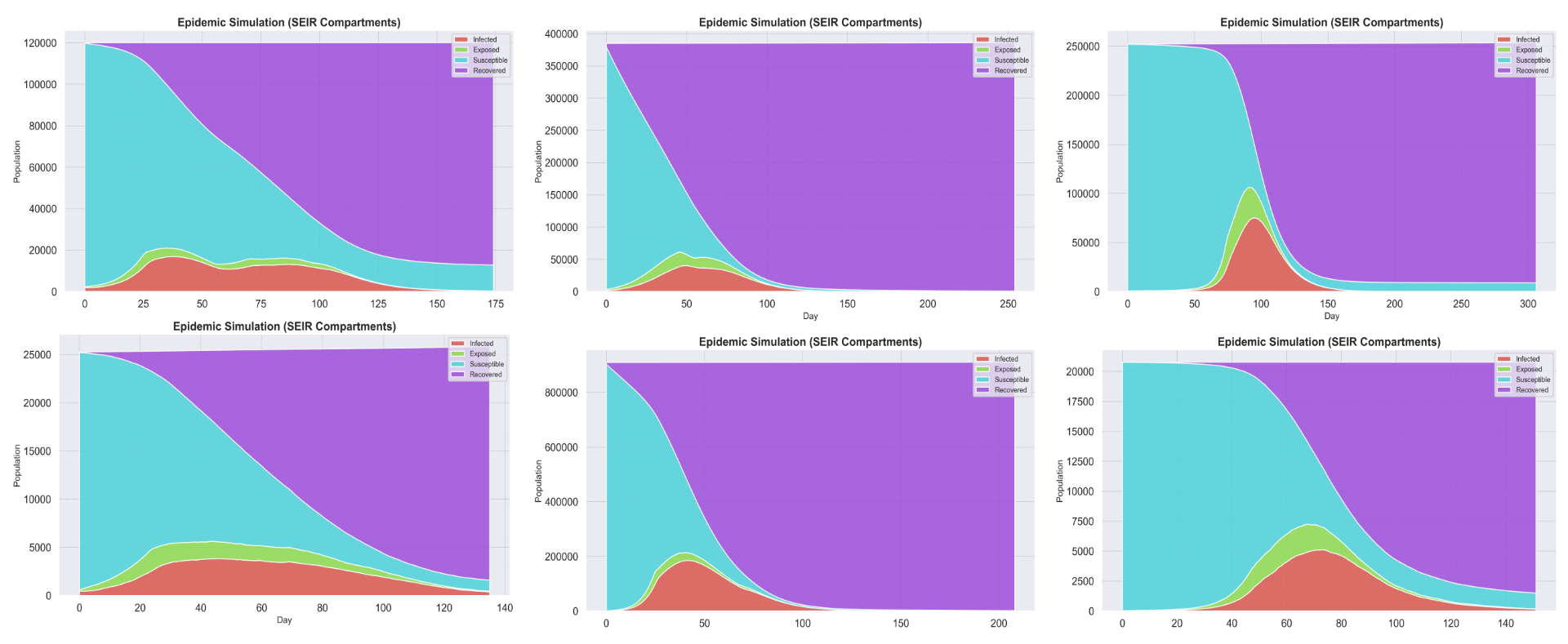} 
    \caption{Batch mode (Python BIGBOY1.2.py batch 6 --plots all)}
    \label{fig:my_figure}
\end{figure}
\subsection{Mathematical Foundations of BIGBOY1.2}
As discussed, BIGBOY1.2 builds upon the foundational SEIR model, which categorizes the population into four compartments \cite{hethcote}. At any point in time, each individual belongs to one of these compartments, and transition between them is governed by a set of differential equations; this has been done to incorporate real-world effects such as vaccination, behavioral factors, and seasonality \cite{bubar2021model, neipel2020seasonal}. \\
\begin{equation} \label{eq:SEIR Fundamental}
\begin{aligned}
    \frac{dS}{dt} &= -\beta(t) \cdot \frac{SI}{N} - \nu S \\
    \frac{dE}{dt} &= \beta(t) \cdot \frac{SI}{N} - \sigma E \\
    \frac{dI}{dt} &= \sigma E - \gamma I \\
    \frac{dR}{dt} &= \gamma I + \nu S
\end{aligned}
\end{equation}

Here is a description of the model's parameters:
\begin{itemize}
    \item $S(t)$, $E(t)$, $I(t)$, and $R(t)$ represent the number of susceptible, exposed, infectious, and recovered individuals at time $t$, respectively.
    \item $N$ is the total population size (assumed constant).
    \item $\beta(t)$ is the time-varying transmission rate, which is crucial for defining how fast susceptible individuals become exposed.
    \item $\sigma$ is the rate at which exposed individuals become infectious ( the inverse of the incubation period).
    \item $\gamma$ is the recovery rate.
    \item $\nu$ is the vaccination rate, which transfers susceptible individuals directly into the recovered class.
\end{itemize}
This extended form of SEIR formulation allows BIGBOY1.2 to simulate epidemic dynamics with the effects of public health interventions such as mass vaccination. \\
Time dependent transmission rate $\beta(t)$ is a powerful and novel feature of BIGBOY1.2. The framework does not assume a constant rate of disease transmission; rather, it models $\beta(t)$ as a function of multiple interacting factors, each of which represents a real world influence on transmission dynamics. 
\begin{equation} \label{eq:beta_t_definition}
\beta(t) = \beta_0 \cdot (1 - \theta_m m(t)) \cdot (1 + \theta_c c(t)) \cdot \left[ 1 + \alpha \sin\left(\frac{2\pi t}{T_s}\right) \right] \cdot \Phi(t)
\end{equation}
Where 
\begin{itemize}
    \item $\beta_0$ is the baseline transmission rate, in the absence of external modifiers.
    \item $m(t)$ is the mask adherence score at time $t$, normalized between 0 and 1.Higher the mask adherence score, higher the compliance with mask wearing. The weight $\theta_m$ controls how strongly this factor suppresses transmission.
    \item $c(t)$ is the crowdedness score, also on a normalized scale. In this, the average density of human interaction is captured, with $\theta_c$ amplifying its effect on transmission.
    \item The sinusoidal term $\alpha \sin\left( \frac{2 \pi t}{T_s} \right)$ models seasonality, it represents periodic increases or decreases in transmission due to environmental and behavioral cycles). $T_s$ is the seasonal cycle period (typically 365 days).
    \item Finally, $\Phi(t)$ is a multi-wave adjustment factor, this allows the simulation to include multiple waves ( due to new variants or changes in social behavior). It is defined as:
\end{itemize}
\begin{equation} \label{eq:phi_t}
\Phi(t) = 1 + \sum_{j=1}^{W} (\phi_j - 1) \cdot \sigma_j(t)
\end{equation}
Here each $\phi_j$ represents the peak multiplier of the $j$-th wave, and $\sigma_j(t)$ is a logistic ramp function that smoothly increases and decreases during the wave period. This component enables multiple waves having sharp rises and slow declines in transmission, a feature often seen in real epidemic data. \\
BIGBOY1.2 supports simulations with heterogeneous population structures, segmented by age and contact environments. This structure is implemented using an age and layer stratified SEIR model. In configurations like this, the population is divided into \textit{ L }contact layers, such as household, workplace, school or community. \textit{A} age groups, such as children, adults and the elders. For each combination the simulation tracks : 
$S_{l,a}$, $E_{l,a}$, $I_{l,a}$, and $R_{l,a}$. \\
Where, \textit{l} = 1,2,...,\textit{L} denotes the contact layer and \textit{a} = 1,2,...,\textit{A} denotes the age group. \\
The force of infection $\lambda_{l,a}(t)$, or the probability per unit time that a susceptible individual in group $(l,a)$ becomes exposed, is calculated using summated contributions from all other groups based on this structured contact matrix \cite{prem2017projecting}:
\begin{equation} \label{eq:force_of_infection}
\lambda_{l,a}(t) = \sum_{l'=1}^{L} \sum_{a'=1}^{A} \beta_{l,a,l',a'}(t) \cdot \frac{I_{l',a'}(t)}{N_{l',a'}}
\end{equation}
This implies that the exposure risk for a school-going kid within the community layer depends on how many infectious individuals exist in other age groups and settings, modulated by the contact matrix $C_{l,l'}$. This method brings realism into the simulation, and modeling of targeted interventions could also be enabled (like school closure or age-prioritized vaccination) \cite{bubar2021model}. \\
The base SEIR model along with above extensions we have done, the BIGBOY1.2 provides a mechanistic ground truth view of an outbreak but real world surveillance data is noisy and subject to various distortions as well. To mimic this effect, BIGBOY1.2 introduces a post processing layer that applies multiple forms of noise and uncertainty to the generated data \cite{gostic2020}. \\
\textbf{Travel Noise} in real epidemics, the geographical boundaries of a population are not sealed. People travel in and out of the region for work, migration and emergencies. The local outbreak curves are affected by this movement, often introducing sudden spikes or dips. We have simulated this behavior through a travel noise generator, which adds or subtracts random infectious cases from the SEIR-generated curve. At each timestep $t$, the infectious compartment $I(t)$ is changed by:
\begin{equation*}
I'(t) = I(t) + \Delta_{\text{travel}}(t)
\end{equation*}
Where $\Delta_{\text{travel}}(t) \sim \mathcal{N}(\mu, \sigma^2)$, a Gaussian-distributed noise term with mean $\mu$ and standard deviation $\sigma$. These parameters can also be fixed by the user. This gives noisy, jagged , heavy-tailed curves that retain the overall trend of the outbreak and also include short-term fluctuations mimicking travel between cities. \\
\textbf{Random Dropper} , another realism challenge in epidemiology is underreporting of cases; all infections are not captured. This may be due to various reasons, maybe because a computer simulation is not really a real-life outbreak scenario. So, to reflect this, we have included a random dropper that hides a certain fraction of cases from the output. 
\begin{equation*}
\text{Reported}_I(t) \sim \text{Binomial}(I'(t), p_r)
\end{equation*}
Where:
\begin{itemize}
    \item $I'(t)$ is the noisy infectious count after travel adjustment.
    \item $p_r \in [0,1]$ is the reporting probability.
\end{itemize}
This same method can be applied to independent exposed, recovered, depending on the use case.\\
We can define the output of BIGBOY1.2 as a function : 
\begin{center} 
\tcbox[
    colback=yellow!5,  
    colframe=black!50, 
    boxrule=0.5pt,    
    arc=0mm, 
    boxsep=5pt     
]
{
$
\mathcal{D}_{\text{BIGBOY1.2}} = \mathcal{R}\left(\mathcal{N}\left(\mathcal{S}_{\text{SEIR}}\left(\mathbf{\Theta}, \beta(t), \Phi(t), \nu, \mathbf{C}, \mathbf{M}, \mathbf{A}\right)\right)\right)
$
}
\end{center}
Where: 
\begin{itemize}
    \item $\mathcal{D}_{\text{BIGBOY1.2}}$: Final reported dataset
    \item $\mathcal{S}_{\text{SEIR}}$:  SEIR simulator that generates compartment curves over time.
    \item $\boldsymbol{\Theta}$: Core epidemiological parameters  $\{\beta_0, \gamma, \sigma, N\}$.
    \item $\beta(t)$: Time varying transmission function.
    \item $\Phi(t)$: Multi wave logistic ramp (captures new waves).
    \item $\nu$: Vaccination rate.
    \item $\mathbf{C}$: Contact matrix.
    \item $\mathbf{M}, \mathbf{A}$: Layer $\mathbf{M}$ and age-group $\mathbf{A}$ structures.
\end{itemize}

Then:
\begin{itemize}
    \item $\mathcal{N}(\cdot)$: Noise layer, which applies:
    \begin{itemize}
        \item Travel noise: $\Delta_{\text{travel}}(t) \sim \mathcal{N}(\mu, \sigma^2)$.
        \item Reporting delay.
        \item Weekend or weekday bias.
    \end{itemize}
    \item $\mathcal{R}(\cdot)$: Reporting layer, which applies:
    \begin{itemize}
        \item Random dropper: $\text{Binomial}(I'(t), p_r)$.
        \item Reporting frequency control (daily, weekly, etc.).
    \end{itemize}
\end{itemize}
The final synthetic dataset $\mathcal{D}_{\text{BIGBOY1.2}}$ is created by first running a SEIR simulation $\mathcal{S}_{\text{SEIR}}$. Then, stochastic noise $\mathcal{N}$ and distortions are injected. Finally, a reporting filter $\mathcal{R}$ simulates real-world underreporting.
\section{Simulation Pipeline}
BIGBOY1.2 is made as a modular simulation engine , it is structured into well-defined functional blocks. Each module processes data through a deterministic or stochastic transformation, allowing control and reproducibility. The system is configured using parameters.json and put together using a python driver script. \\
\textbf{Configuration Parsing and Preprocessing} : At runtime, the simulation parses a structured parameter file that contains: 
\begin{center}
\begin{tcblisting}{
  listing only,
  hbox,
  colback=blue!5,
  colframe=black!50,
  boxrule=0.5pt,
  arc=0mm,
  boxsep=5pt,
  listing options={style=jsonstyle}
}
{
    "population": 20785,
    "days": 152,
    "initial_infected": 32,
    "mask_score": 10,
    "crowdedness_score": 7,
    "quarantine_enabled": "y",
    "seasonality_enabled": "y",
    "interventions_enabled": "n",
    "reporting_prob_min": 0.52,
    "reporting_prob_max": 0.72,
    "multi_wave": "n",
    "random_seed": 845114,
    "vaccination_enabled": "n",
    "daily_vaccination_rate": 0.016,
    "incubation_period": 5,
    "waves": [
        {
            "day": 60,
            "beta": 2.5,
            "seed": 100
        }
    ],
    "testing_rate": "medium",
    "mask_decay_rate": 0.0156,
    "travel_enabled": "n",
    "travel_max": 0,
    "mode": "random",
    "layers": 2,
    "age_groups": 3
}
\end{tcblisting}
\end{center}
Above is a sample params.json taken from a generated batch. The parser validates all input types, auto generates required time series and prepares input buffers for the simulation. \\
\textbf{Compartmental Simulation Layer}: This module numerically integrates a layered, age structured SEIR system. It implements forward Euler integration over discrete time stamps, contains $L \times A$ compartment states in 4D tensors.
\begin{center}
\tcbox[
  colback=blue!5,
  colframe=black!50,
  boxrule=0.5pt,
  arc=0mm,
  boxsep=5pt
]{
  $S[L, A], \quad E[L, A], \quad I[L, A], \quad R[L, A]$
}
\end{center}
Transmission rate $\beta_t$ is calculated per time stamp by combining time dependent behavioral scores (mask , crowd), seasonal effects (sinusoidal) and wave ramp function. Cumulative states are kept in the memory and this engine supports toggling between homogenenous and stratified contact modes.\\
\textbf{Multiwave Modulation}, this submodule applies wave-shaped multipliers on beta effective. \textbf{Noise Injection Module}, wraps the raw SEIR outputs and introduces realistic distortions such as travel noise, delay, random modulators, and zero clipping.\\
\textbf{Output and Export Handlers} are responsible for making time series CSVs for reported date , it includes 2 CSV files, one containing just the reported cases and the other containing: Day, Susceptible, Exposed, Infected, Recovered, New Exposed, New Infections, New Recoveries, Reported Cases,	$\beta_t$, Seasonality, $R_t$. Output manager is also responsible for diagnostic logs (parameter hash, seed) and optional visualizations.
\begin{figure}[H] 
    \centering
    \includegraphics[width=0.8\linewidth]{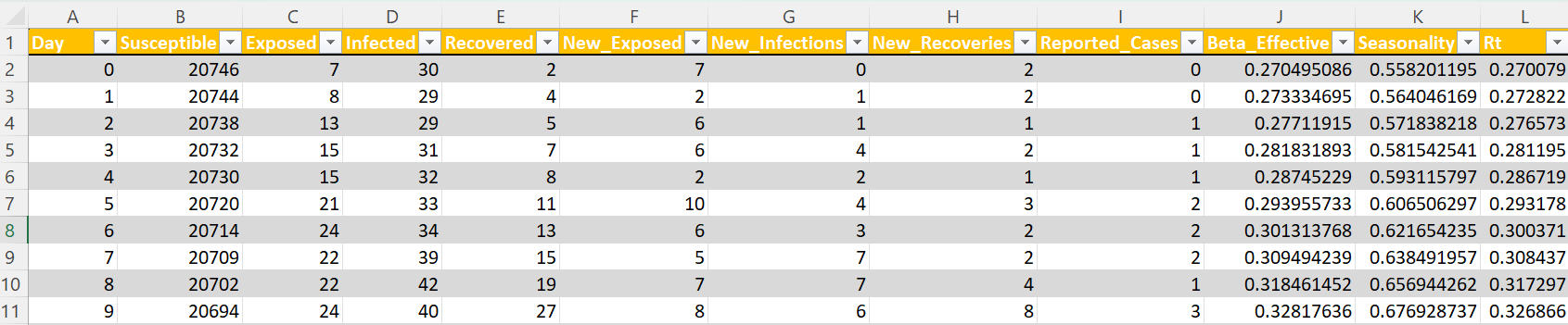} 
    \caption{Dataset Snapshot from a random batch}
    \label{fig:my_figure}
\end{figure}
\begin{figure}[H] 
    \centering
    \includegraphics[width=0.8\linewidth]{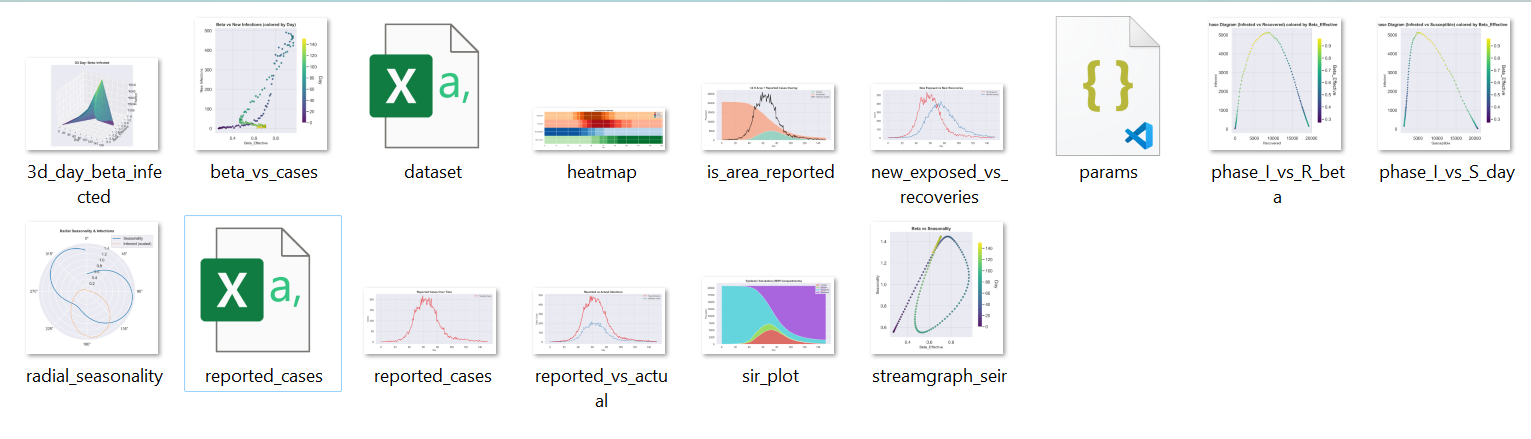} 
    \caption{Directory snapshot of save CSVs, JSON and PNGs}
    \label{fig:my_figure}
\end{figure}
\textbf{Reproducibility and Logging}: reproducibility is done by random seeds, this allows any experiment to be fully replicated or benchmarked until the user hasn't deleted the params.json file.
\section{Visual Plots and Demonstration}
Visual Plots are an integral part of BIGBOY1.2, as they present epidemic simulation data in interpretable and high-dimensional plots. These visual plots could be used as analytical tools that could validate model outputs and also reveal deeper insights like epidemic progression, interventions, and control.
\subsection{Overview of the Plotting System}
Upon simulation, BIGBOY1.2 allows the users to generate a diverse set of plots by passing --plots all or --plots sir. It has a modular plotting engine that allows both minimal and advanced visualization, and the output is saved as high resolution PNG files. All of the plots are further saved in timestamped directories with accompanying metadata, ensuring reproducibility.
\subsection{SEIR Compartment Plot}
A stacked area plot showing the progression of Susceptible (S), Exposed (E), Infected (I) and Recovered (R) populations over time, classical SEIR-style visualization is foundational for understanding the macro level behavior of the epidemic. The purpose of the SEIR-stacked chart is revealing key phases such as exponential growth, peak infections and herd immunity thresholds. 
\begin{figure}[H]
    \begin{minipage}{0.49\linewidth}
        \flushleft
        \includegraphics[width=\linewidth]{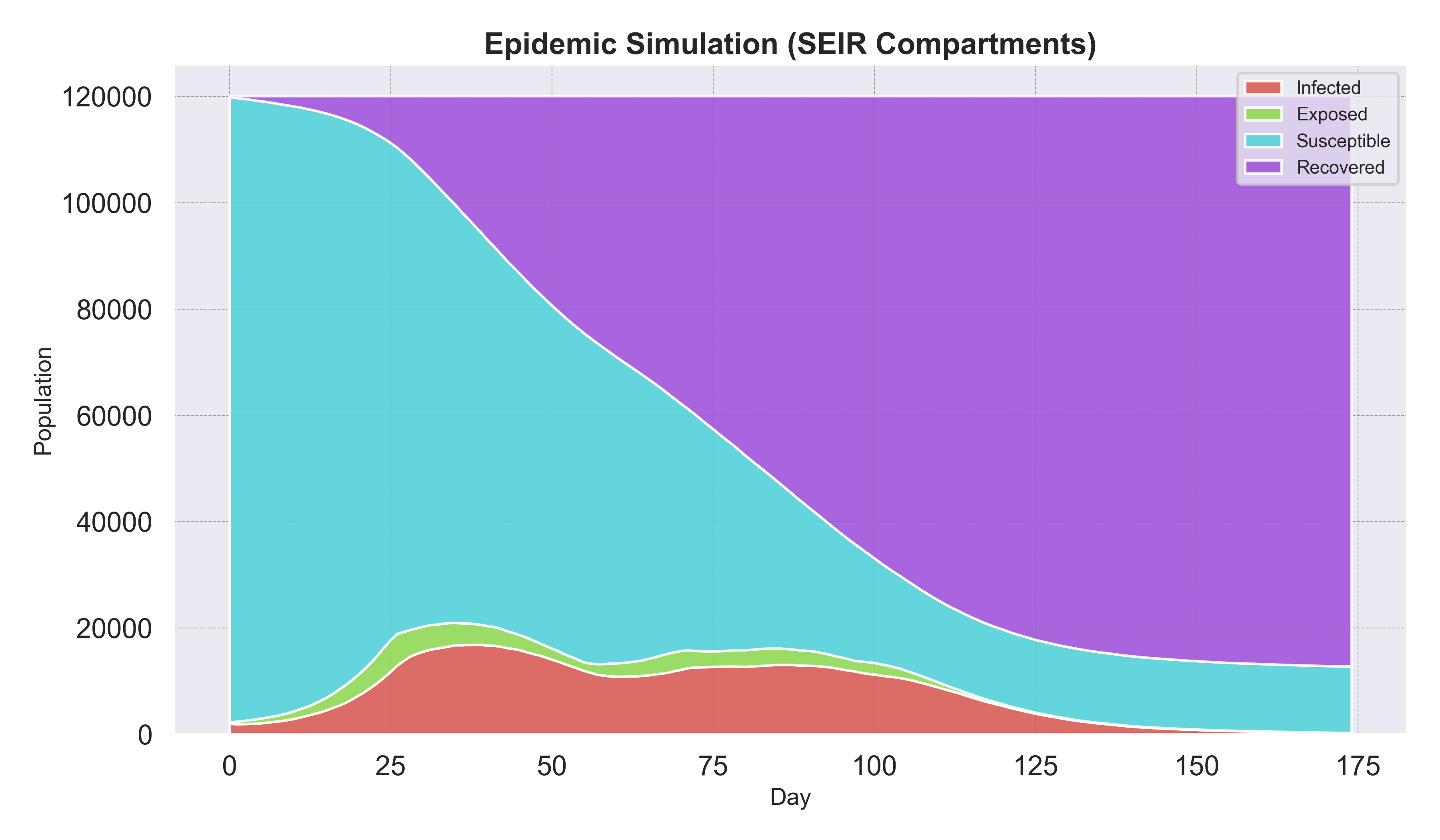}
    \end{minipage}
    \hfill
    \begin{minipage}{0.49\linewidth}
        \flushright
        \includegraphics[width=\linewidth]{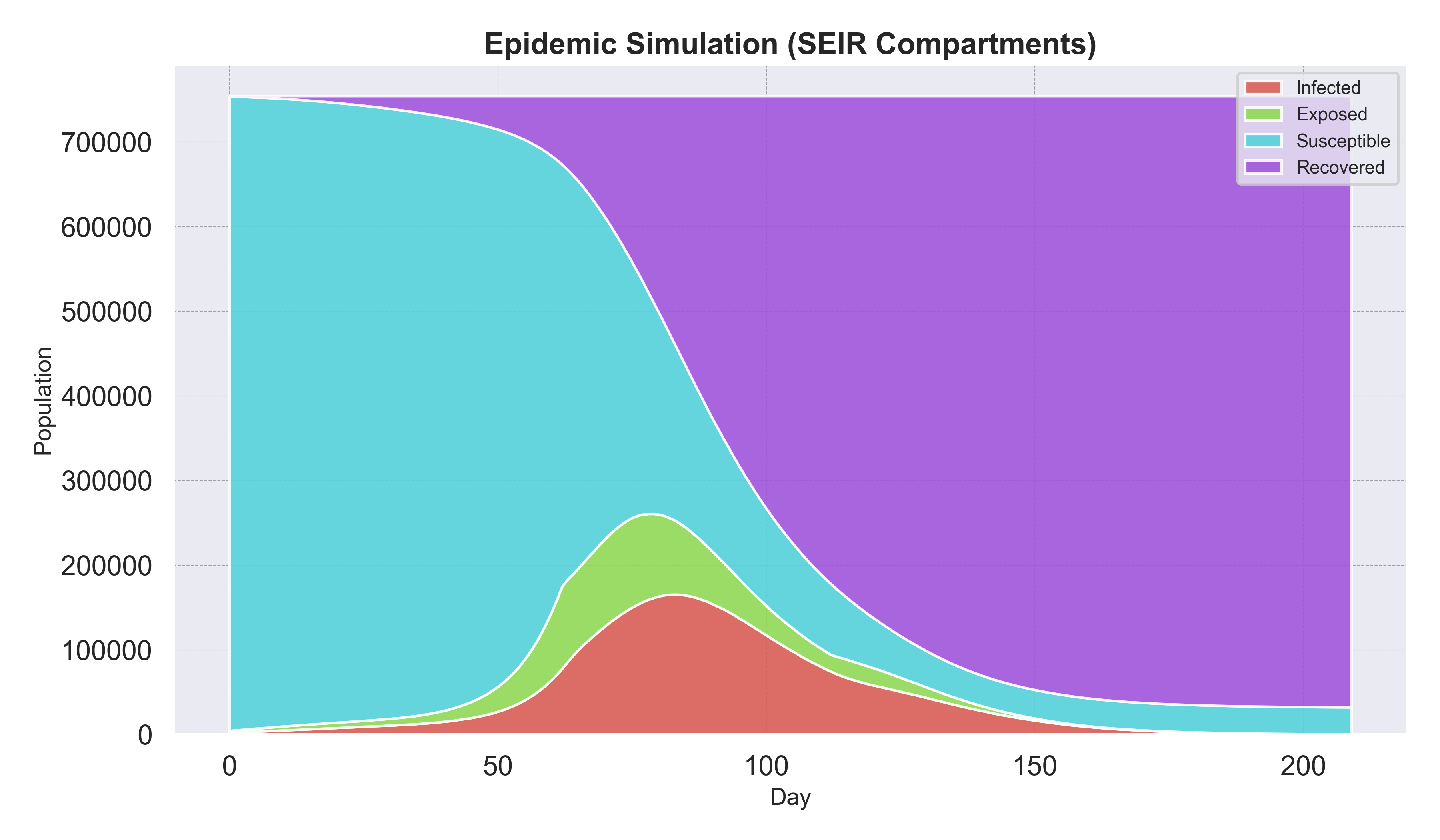}
    \end{minipage}
    
    \caption{SEIR stacked plots from a random batch}
    \label{fig:seir-side-by-side}
\end{figure}

\subsection{Reported Cases Timeline}
It displays the reported cases across days; this contrasts the real-world observed data with latent epidemic dynamics. It simulated the public health reporting pattern .
\begin{figure}[H]
    \begin{minipage}{0.49\linewidth}
        \flushleft
        \includegraphics[width=\linewidth]{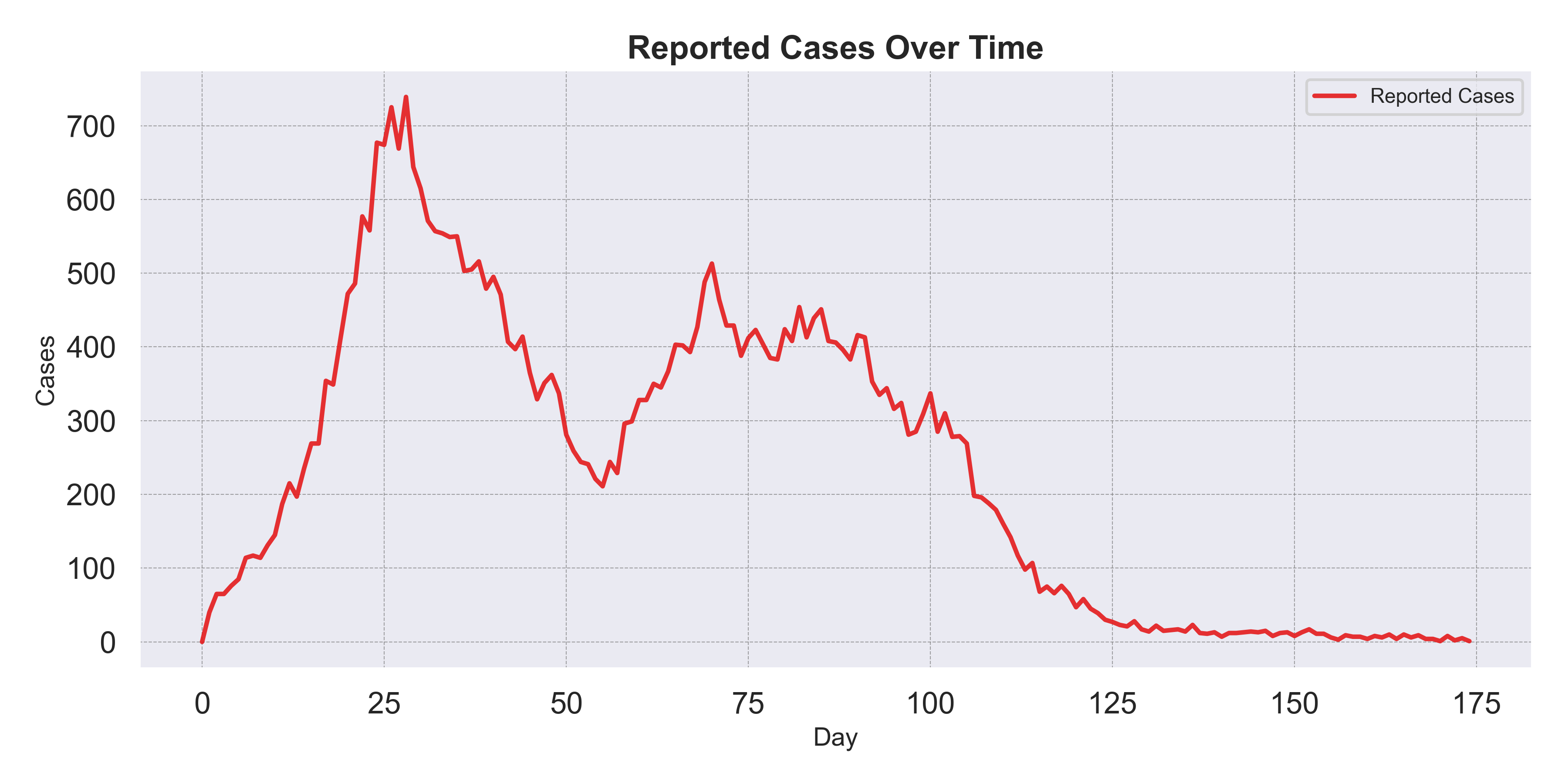}
    \end{minipage}
    \hfill
    \begin{minipage}{0.49\linewidth}
        \flushright
        \includegraphics[width=\linewidth]{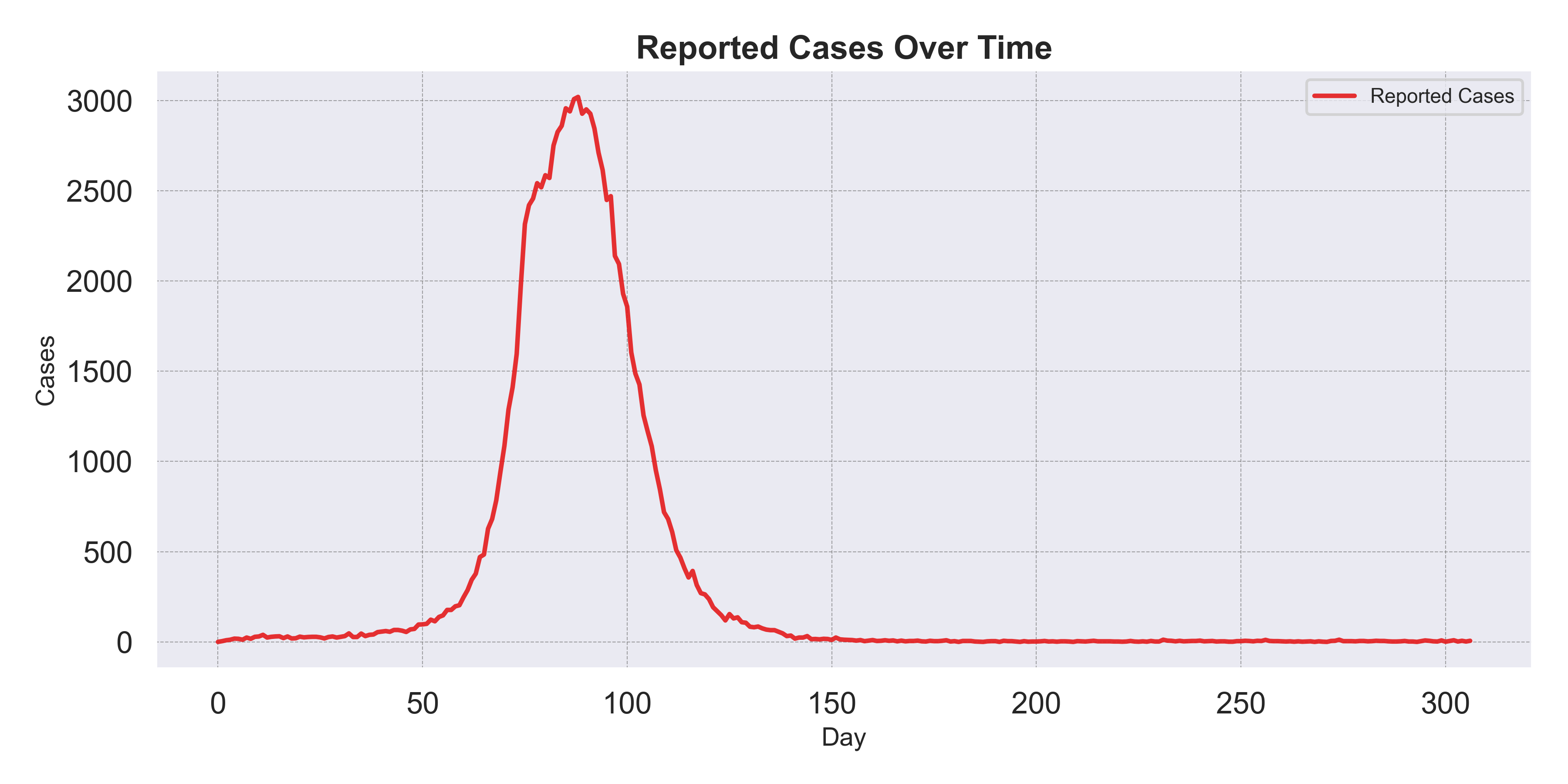}
    \end{minipage}
    
    \caption{Reported cases from a random batch}
    \label{fig:seir-side-by-side}
\end{figure}

\subsection{3D plot of day $\times \beta \times$ infection}

This 3D plot displays how contagious a disease is, as contingency and number of infections are not linear. The 3D visualization shows how small changes in $\beta$ can lead to explosive outbreaks under certain conditions. Lag effects can also be revealed by visualizing this 3D plot. Even if $\beta$ rises sharply, infections may spike a few days later; this helps in understanding incubation periods. \\
If an intervention like mask adherence is applied (or lockdown) , it causes $\beta$ to drop and the flattening infection counts could be seen in the Z-axis. This shows causal evidence of policy effectiveness in time. In multiple wave simulations, the 3D plot clearly shows how subsequent waves differ in timing, strength and transmissibility. These could be compared early vs later variants of the disease visually. 
\begin{figure}[H]
    \begin{minipage}{0.49\linewidth}
        \flushleft
        \includegraphics[width=\linewidth]{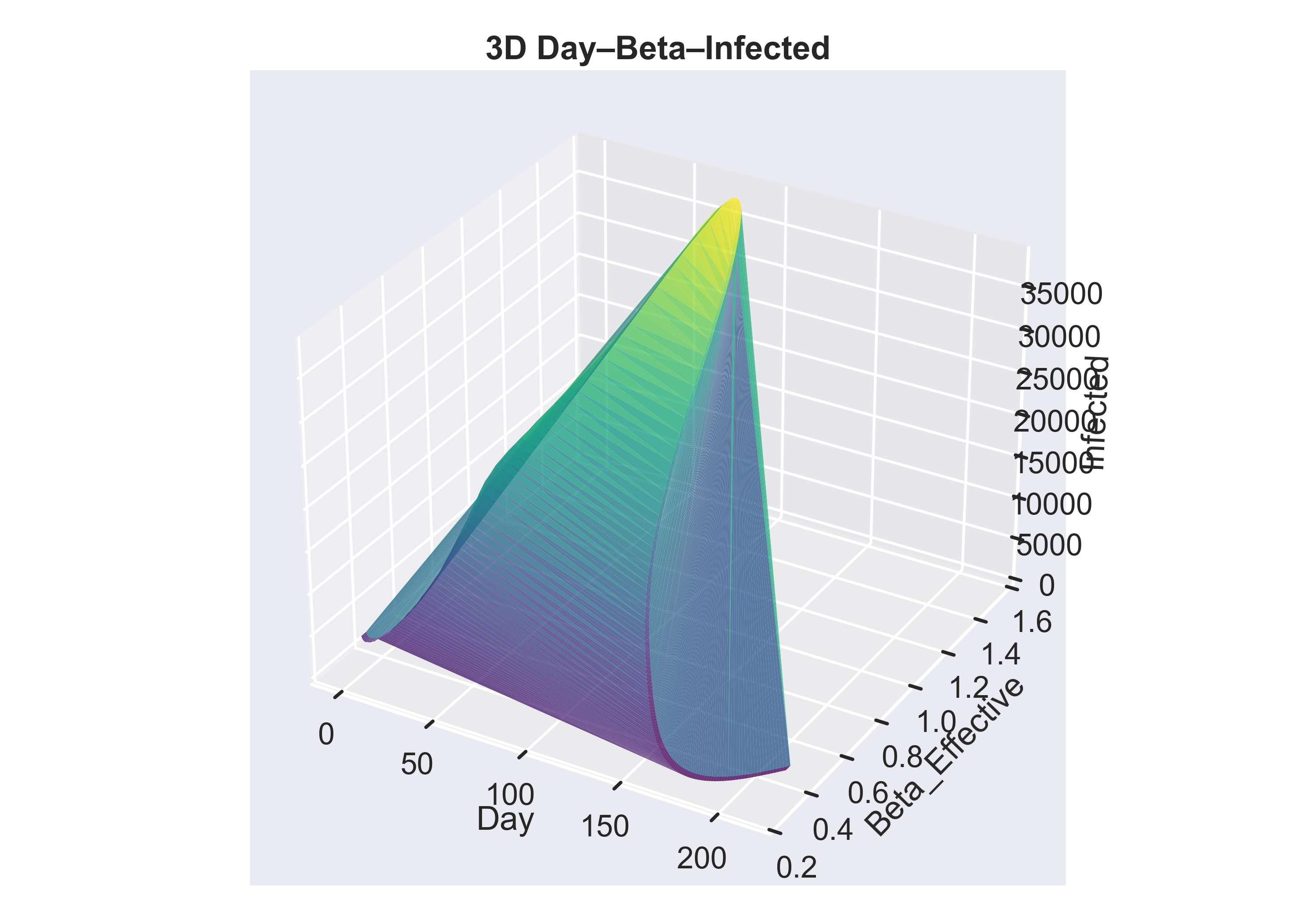}
    \end{minipage}
    \hfill
    \begin{minipage}{0.49\linewidth}
        \flushright
        \includegraphics[width=\linewidth]{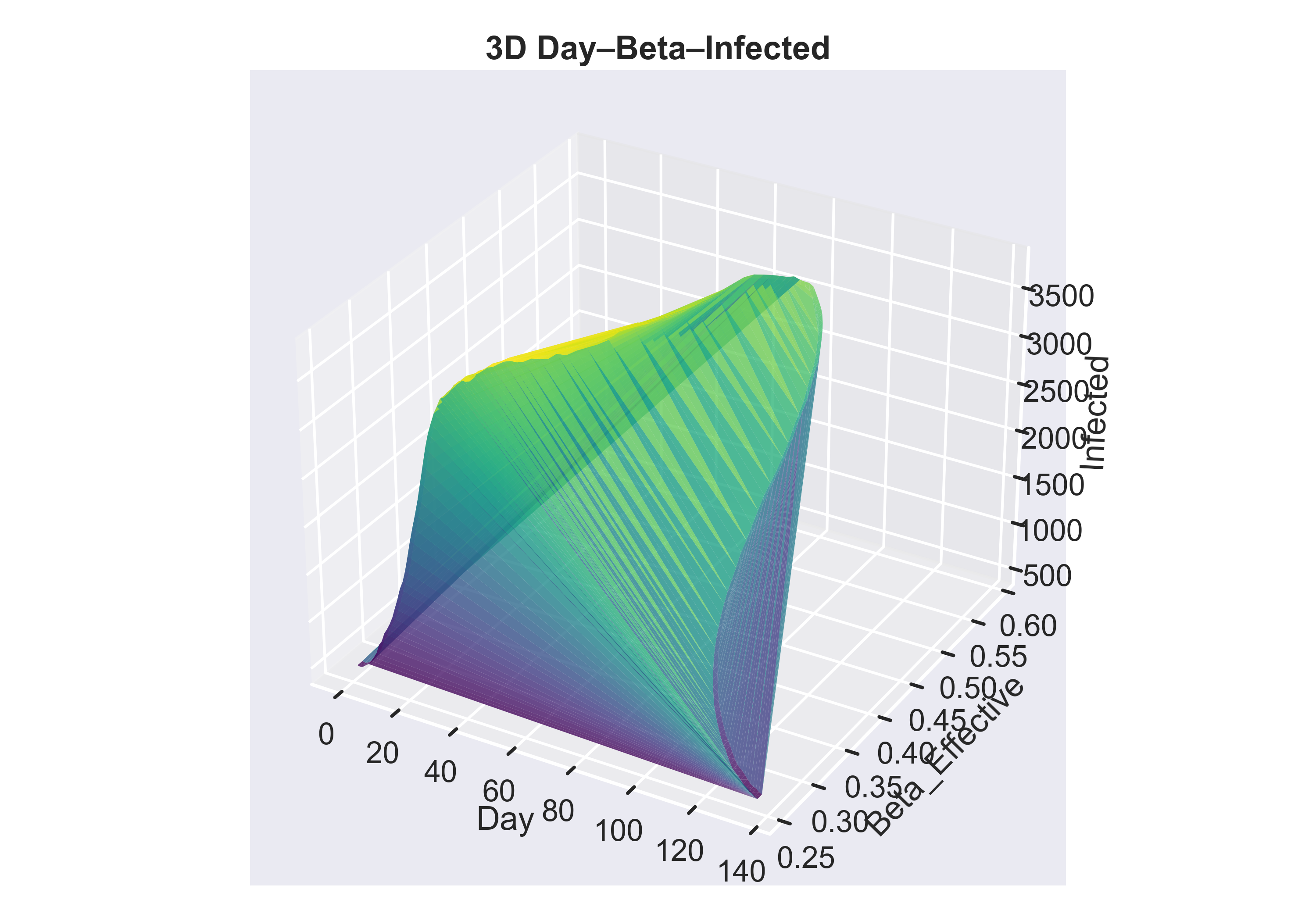}
    \end{minipage}
    
    \caption{3D plots from a random batch}
    \label{fig:seir-side-by-side}
\end{figure}
\subsection{$\beta$ vs. New Infections (Colored by Day)}
This scatterplot is crucial for understanding how changes in $\beta$ correlate with spikes or drops in new infections. Higher $\beta$ generally increases the infections, but during later stages, even higher $\beta$ may not cause spikes due to immunity build up. These effects are shown by Day, the scatter dots are colored by day.
\begin{figure}[H]
    \begin{minipage}{0.49\linewidth}
        \flushleft
        \includegraphics[width=\linewidth]{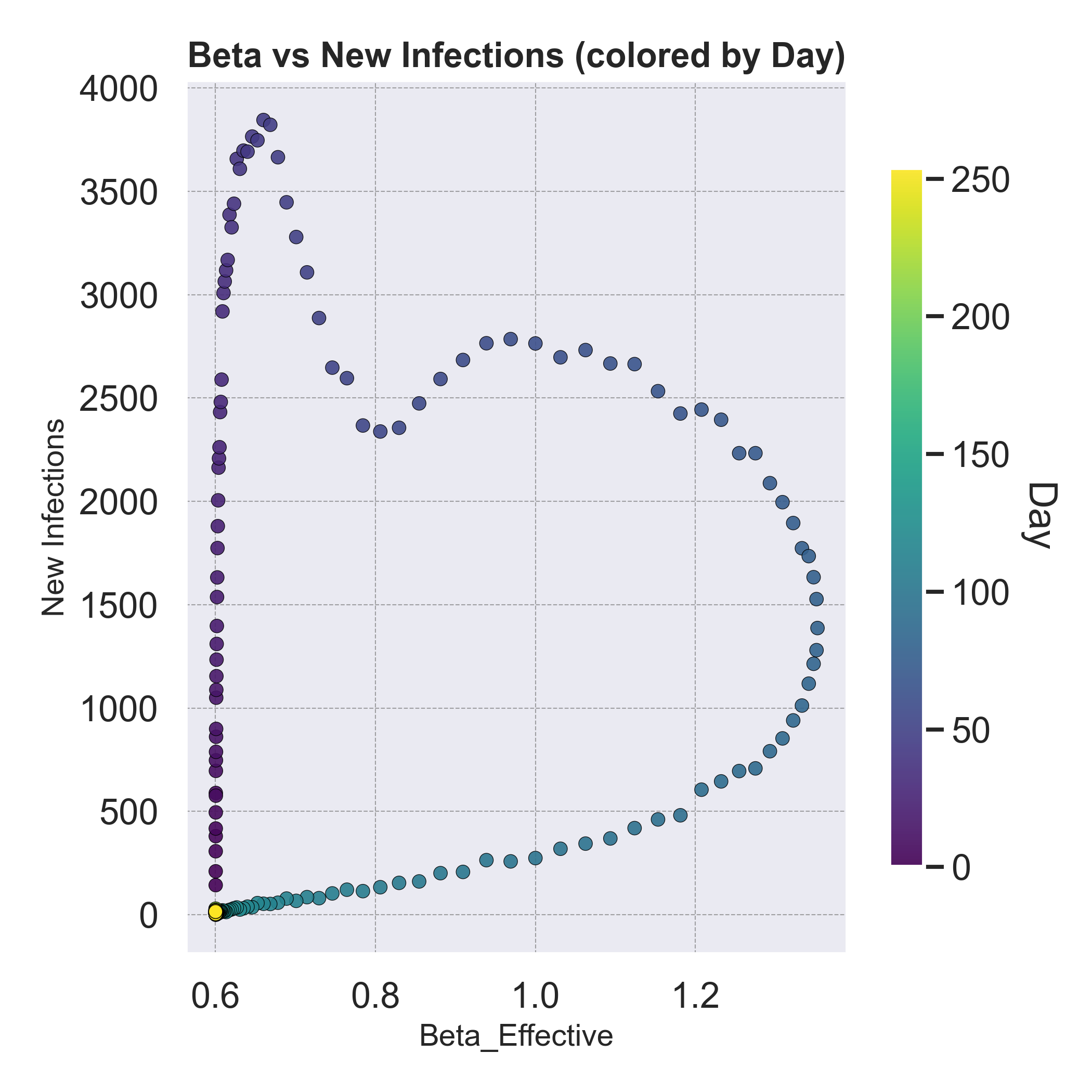}
    \end{minipage}
    \hfill
    \begin{minipage}{0.49\linewidth}
        \flushright
        \includegraphics[width=\linewidth]{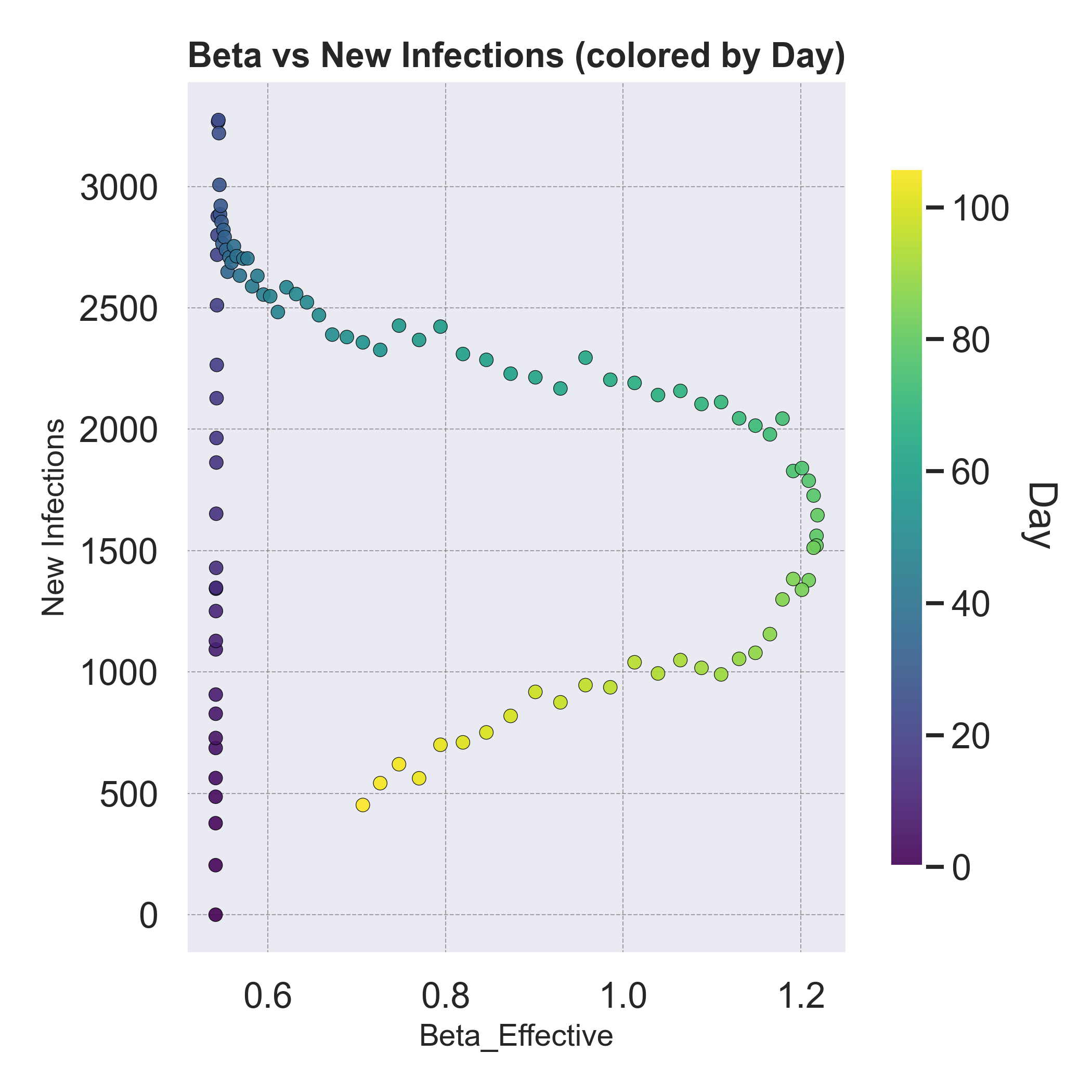}
    \end{minipage}
    
    \caption{Beta vs. New Infections plot from a random batch}
    \label{fig:seir-side-by-side}
\end{figure}
\subsection{New Exposed vs New Recoveries}
This plot helps in epidemic growth detection. If New exposed > New Recoveries, the infection is spreading faster than it's being cleared, and if New exposed < New Recoveries, it indicates a decline in the epidemic. \\
This plot often shows a crossover point where the two curves intersect each other. The first crossover signals the wave onset and the second crossover suggests wave resolution or success in interventions. \\
Reduced transmission is demonstrated by sharp dips in new exposures after a policy change, for example, lockdown and vaccination. If new recoveries continue to rise , it means that the healthcare system is still managing the previous cases, but future burden is dropping.
\begin{figure}[H]
    \begin{minipage}{0.49\linewidth}
        \flushleft
        \includegraphics[width=\linewidth]{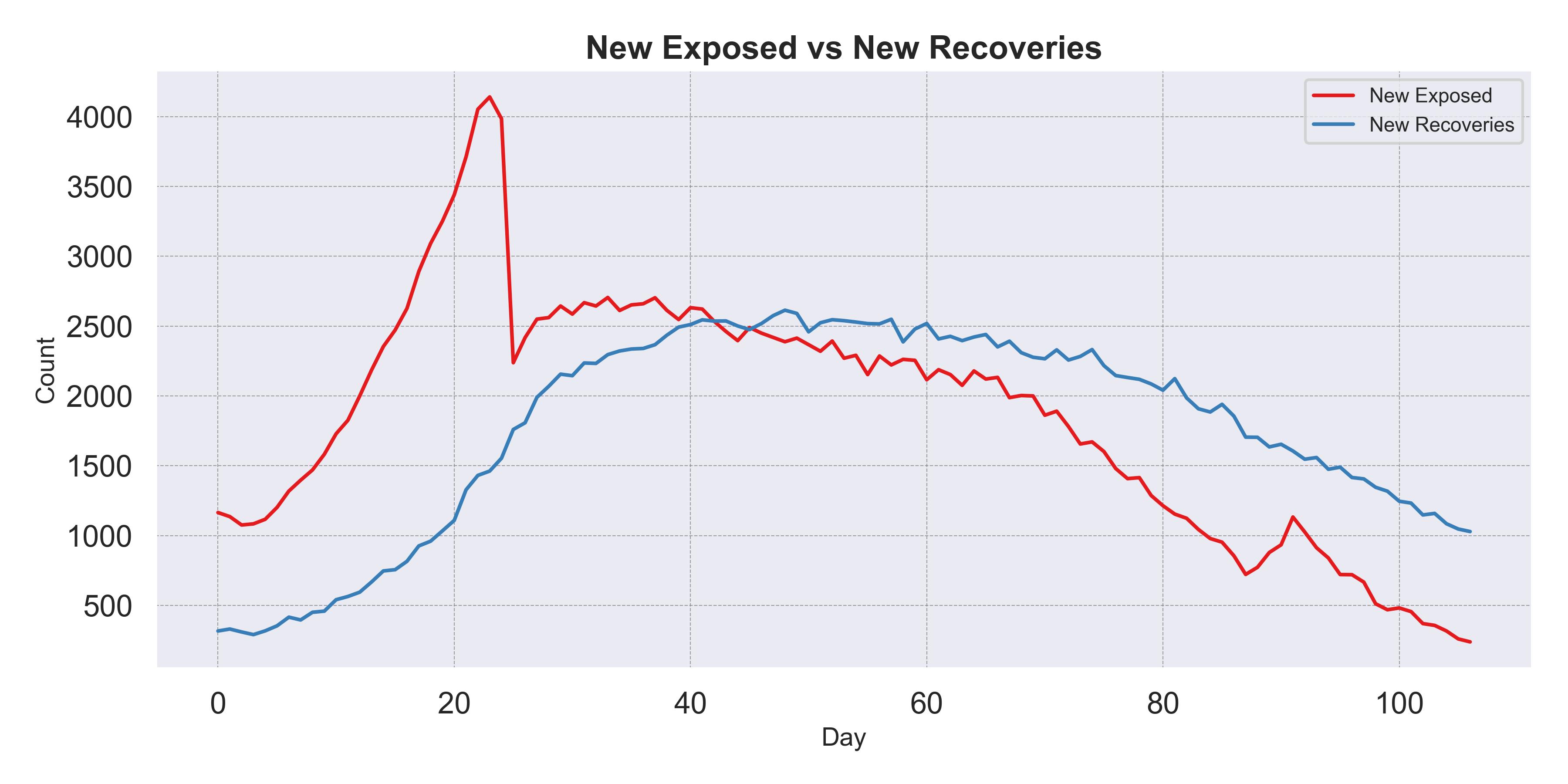}
    \end{minipage}
    \hfill
    \begin{minipage}{0.49\linewidth}
        \flushright
        \includegraphics[width=\linewidth]{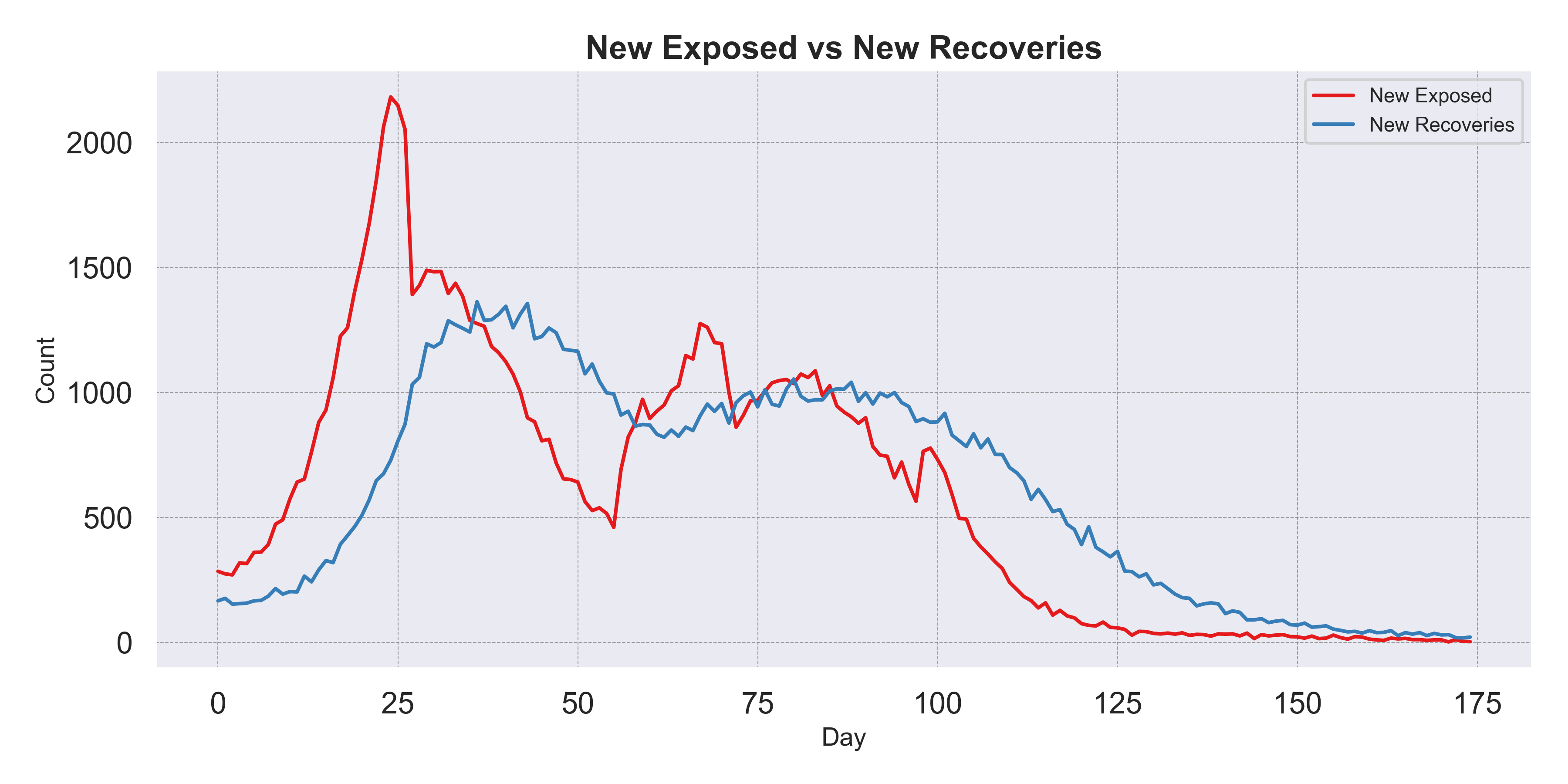}
    \end{minipage}
    
    \caption{New Exposed vs New Recoveries plot from a random batch}
    \label{fig:seir-side-by-side}
\end{figure}
\subsection{Phase Diagrams}
Two types of phase diagrams are displayed, the I vs S diagram and the I vs R diagram, both colored by $\beta$ effective. In the I vs S phase plot, each point represents the system's state on a given day. As the epidemic goes on, the system traces a trajectory through its space, forming a loop or an arc, highlighting the rise and fall of infections relating to the shrinking susceptible population.\\
The I vs R variant similarly tracks how infections transition into recoveries and it is particularly useful for visualizing the cumulative impact of the epidemic. Unlike simpler plots, phase diagrams capture the entire system's evolution in a compact , geometric path , making them crucial for both theoretical exploration and practical modeling.
\begin{figure}[H]
    \begin{minipage}{0.49\linewidth}
        \flushleft
        \includegraphics[width=\linewidth]{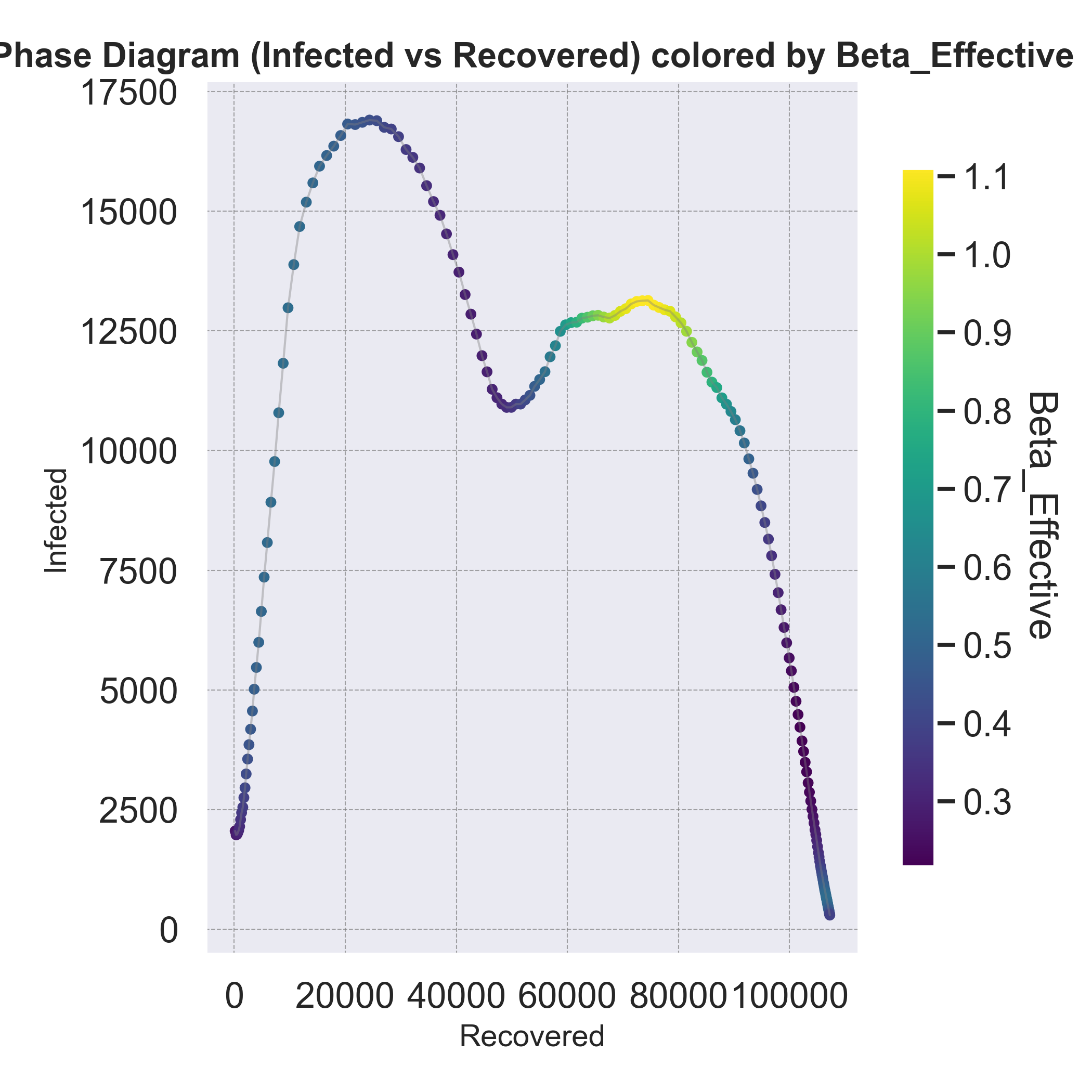}
    \end{minipage}
    \hfill
    \begin{minipage}{0.49\linewidth}
        \flushright
        \includegraphics[width=\linewidth]{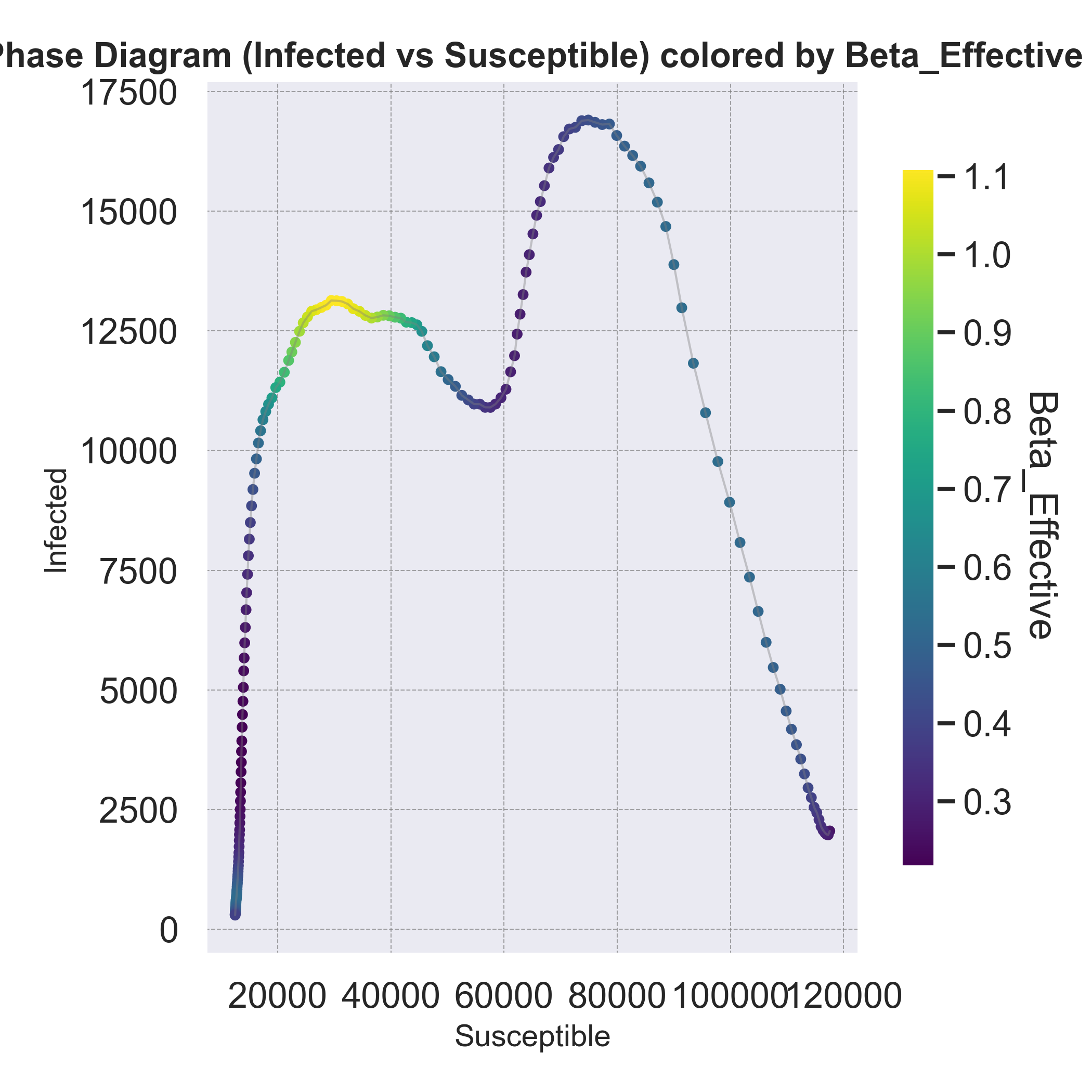}
    \end{minipage}
    
    \caption{Phase Diagrams from a random batch}
    \label{fig:seir-side-by-side}
\end{figure}
\subsection{Othe plots}
The \textbf{radial seasonality} plot uses a polar coordinate system to display cycles of seasons alongside infection levels, which makes it ideal for visualizing periodic disease where the transmission rate is directly related to seasonality. The radial axis shows both seasonality strength and infection counts, which allows the user to correlate peaks in transmissibility with actual outbreaks. \\
The \textbf{Reported vs Actual Infections} is a dual line graph that compares reported cases with new infections which helps users to visualize underreporting, testing delays, and observational noise. It is key for assessing the gap between observed data and true disease data , especially in low testing regions and during surges. \\
\begin{figure}[H]
    \begin{minipage}{0.49\linewidth}
        \flushleft
        \includegraphics[width=\linewidth]{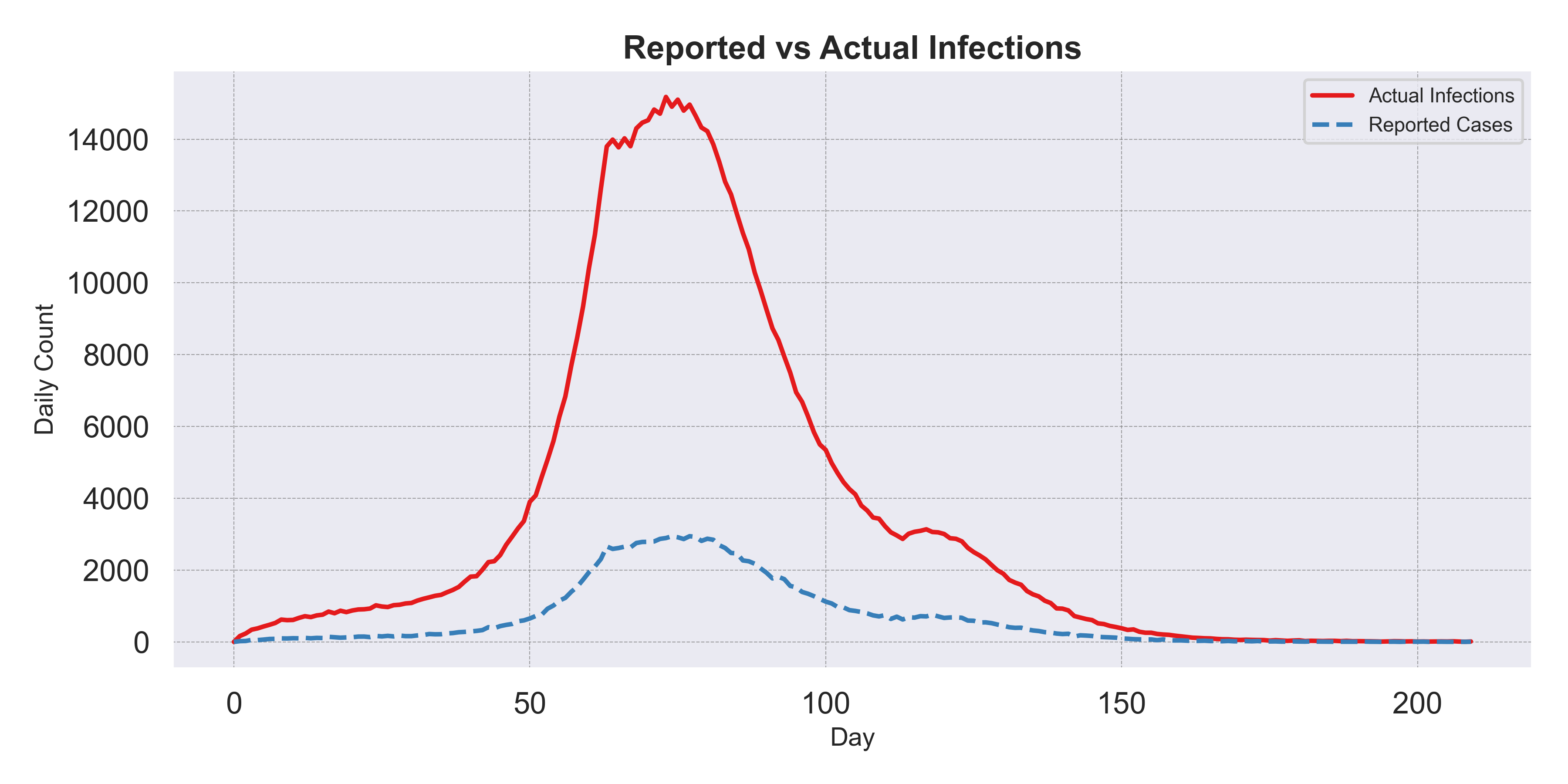}
    \end{minipage}
    \hfill
    \begin{minipage}{0.49\linewidth}
        \flushright
        \includegraphics[width=\linewidth]{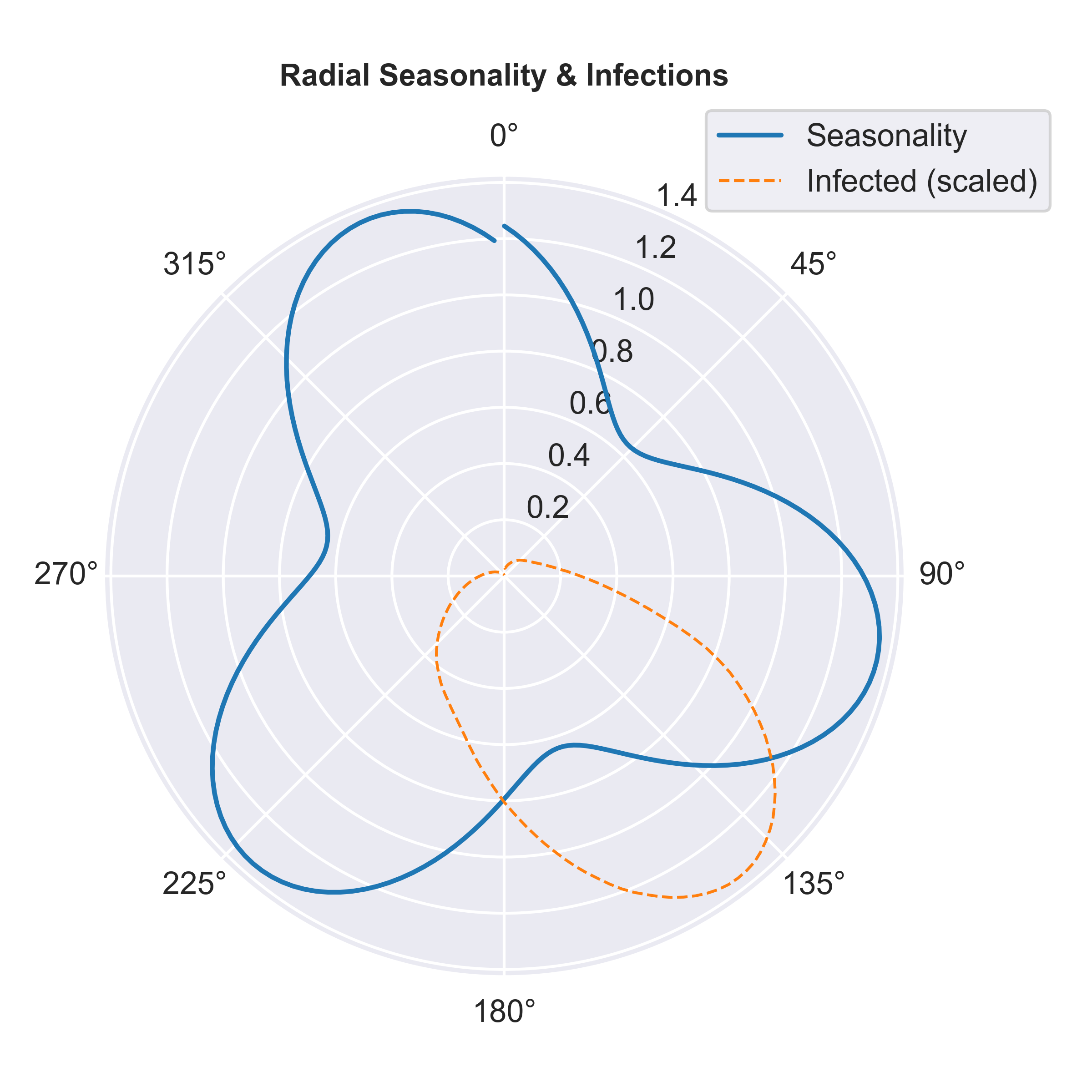}
    \end{minipage}
    
    \caption{}
    \label{fig:seir-side-by-side}
\end{figure}
\subsection{Comparison with real world date}
To evaluate the realism and validity of the synthetic data generated by our model, we have compared it with the actual daily reported cases of COVID-19 in India (for the second wave).The data were taken from the official WHO website \cite{who2021india}, first converted to BIGBOY format (it had only one column of reported cases) using a Python script (available on GitHub), and the comparison was run using another Python script. We used scaling to match the range and resolution of synthetic outputs. The overall epidemic curve shape, peak structure and rise-fall dynamics have remained remarkably consistent across both datasets, despite manual feeding of parameters to BIGBOY1.2. We have done metric based comparisons between the two data below.\\
\begin{figure}[H]
    \centering
    \includegraphics[width=1\linewidth]{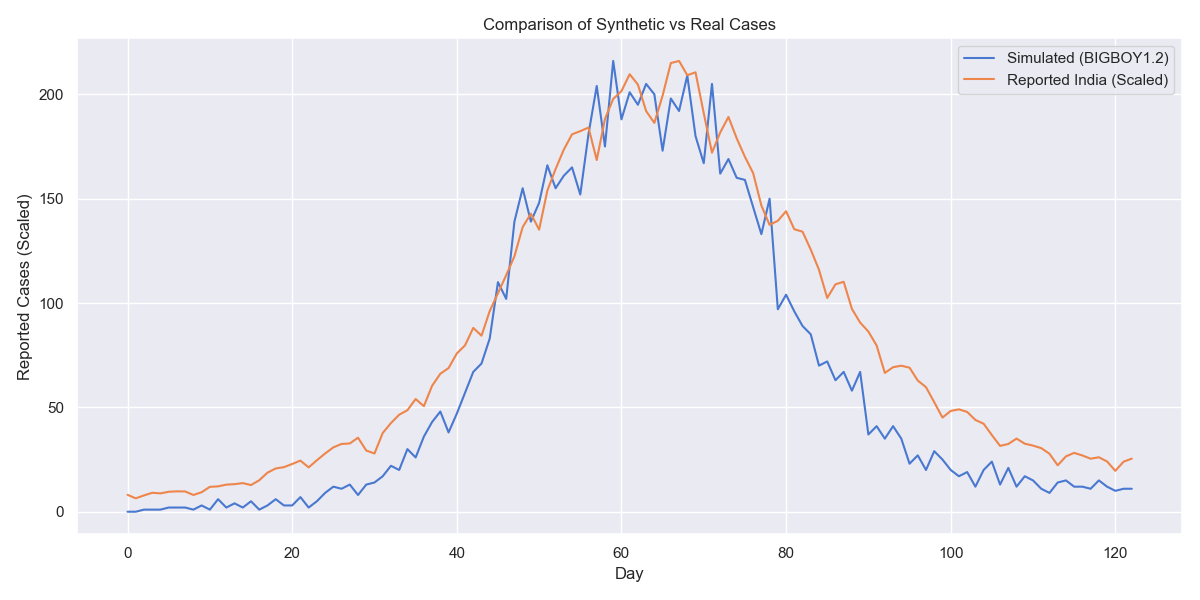}
    \caption{COVID19 2nd wave compared to BIGBOY1.2}
    \label{fig:enter-label}
\end{figure}
\begin{table}[ht]
\centering

\arrayrulecolor{midgray}           
\setlength{\arrayrulewidth}{0.5pt}  

\rowcolors{1}{white}{lightgray}   

\begin{tabular}{|>{\columncolor{lightgray}}l|c|c|}
\hline
\textbf{Metric} & \textbf{Real Data} & \textbf{BIGBOY1.2 Simulated} \\
\hline
Basic Reproduction Number ($R_0$) & 1.29 & 1.16 \\
Epidemic Duration (Days) & 53 to 71 & 51 to 70 \\
Peak Infection Days (Top 3) & 67, 68, 59 & 59, 68, 67 \\
\hline

\end{tabular}

\arrayrulecolor{black}  
\caption{Comparison of COVID-19 vs BIGBOY1.2}
\end{table}
\subsection{Sources of Deviation}
The shape of the waves depicts promising overlap, there are still scopes of improvement, these deviations are not unexpected and can be attributed to several factors like \textbf{real world data complexity} , publicly available data suffers from factors like inconsistent testing , region specific anomalies. Extracting granular data would improve graph alignment more. Another factor is \textbf{parameter calibration}, BIGBOY1.2 uses manually selected parameters, which is good for understanding curves and real world factors but when it comes to mimicking real world curves, automated hyperparameter tuning ( like Bayesian optimization) would match real epidemics more precisely. These changes would be incorporated in the next version of BIGBOY.\\
These minor discrepancies between the simulated and observed curves do not stem from model inaccuracies; rather, it is the inherent stochasticity and undetermined variables that play in the real world outbreaks. In fact, real world epidemics are so chaotic that even the same virus would behave differently if replayed in the same conditions. 

\section{Future Work (BIGBOY1.3)}
BIGBOY1.2 provides a great platform for generating synthetic disease outbreak data, but its development is not completed yet. We have planned to include several powerful extensions for future versions; the roadmap includes the following improvements : 
\subsection{Enhanced Visualizations}
We aim to include an animate mode as --animate X (X being different animated visuals) to make outbreak visualizations more dynamic and interactive. This mode would include layered epidemic progression, animated transmission waves, and geospatial spread mapping. We would also be integrating agent-based and grid-based simulations to complement the compartmental SEIR model. This will allow us to see an individual level perspective, mobility, and stochasticity; it would help in capturing phenomena like superspreading and localized interventions.
\subsection{Disease mode}
A new interface would be implemented in the CLI, which would allow the user to select from pre-configured templates for diseases like COVID-19, measles, influenza, and more. Each of these templates would include pre-loaded parameters, allowing faster and disease-specific scenario generation and modeling. The mode could be accessed as --disease X (X being the disease template). We would also expand on the intervention parameters, and healthcare system constraints would also be implemented. A better multi-wave structure would also be configured.
\subsection{Automated Hyperparameter Tuning}
BIGBOY1.3 will feature automated hyperparameter tuning using techniques such as Bayesian Optimization and grid search \cite{snoek2012practical}, which will calibrate parameters directly from real datasets, which could be used in scenarios where a limited amount of data is available for a particular disease and BIGBOY1.3 could be used to generate unlimited disease-like data using AHT. 
\subsection{Country mode}
BIGBOY1.3 will also feature a country mode where users could select from all the countries and a set of calibrated parameters would be applied to them. These parameters would include population, crowding, literacy (corresponding to mask adherence and vaccinations), country specific behaviors, population pyramids (dividing population into categories) and seasonality profiles. \\
Let us understand capabilities of BIGBOY1.3 using an example.
\begin{center}
\tcbox[
  colback=blue!5,
  colframe=black!50,
  boxrule=0.5pt,
  arc=0mm,
  boxsep=5pt
]{
 --disease EBOLA --country INDIA --state HARYANA --animate GRID
}
\end{center}
This would simulate the outbreak of the EBOLA disease in Haryana, India. Although EBOLA has never hit Haryana, the model contains both the profiles for EBOLA and Haryana and will flawlessly simulate and generate datasets for the Ebola outbreak in Haryana.
\subsection{Ramen1}
We are working on several SVR, SIR, SEIR and FDE \cite{duan2020svr,li2020fractional} based models and their hybrids to make SOUP S-1 (SVR-SIR) , CRUM1 (SVR-SEIR), Broth-N (Naive baseline-based SVR) and SVR-FDE models; furthermore, this model soup would be merged into a high fidelity ensemble system named Ramen1, which would combine the best of all approaches using weighted voting and time series. 
\section{Open Source and Contributions}
BIGBOY1.2 is open source and could be used for unrestricted academic and noncommercial use. The complete codebase along with sample datasets from BIGBOY1, BIGBOY1.1 and BIGBOY1.2 are publicly available on \href{https://github.com/RaunakNarwal735/BIGBOY1.1}{GitHub}. \\
A short demo video demonstrating how to run BIGBOY1.2 on your local machine is available on \href{https://www.youtube.com/watch?v=jOsmuO00low}{Youtube}. \\
We encourage students and researchers to use BIGBOY1.2 to understand and simulate epidemic curves, and we are open to contributors who would like to be a part of BIGBOY1.3 and further improvements. For contribution purposes, mail \href{mailto:ms23177@iisermohali.ac.in}{here}. \\
Ramen1 is the bigger project which required generation of a synthetic dataset, that is how BIGBOY1 came into being. Ramen1 used many sub models to predict disease outbreaks, it switches between various models depending on the phase of the outbreak (ensemble model). It is still a work in progress,  if anyone wants to contribute to Ramen1 , send a mail \href{mailto:ms23177@iisermohali.ac.in}{here}. 

\subsection{Acknowledgements: } The author (Raunak Narwal) would like to thank \href{mailto:abbas@iitmandi.ac.in}{\textbf{Prof. Syed Abbas} }for this internship opportunity which led to this project.\\
The author would also like to thank \href{mailto:Rishunarwal24@gmail.com}{ \textbf{Rishu Narwal}}, Phd candidate at IIT Delhi, for  helping out with research literature, structure of the research paper and providing a computational support.\\
Special thanks to \href{mailto:ms23089@iisermohali.ac.in}{\textbf{Chirag Verma}}, third year undergraduate student at IISER Mohali, for his theoretical insights and discussions in the area of epidemic outbreak modeling.

\bibliographystyle{unsrt} 
\bibliography{references}  

\end{document}